\documentclass[aps,prd,twocolumn,floatfix]{revtex4} 
\usepackage{epsfig} 
 
\def\gsim{ \lower .75ex \hbox{$\sim$} \llap{\raise .27ex \hbox{$>$}} }  
\def\lsim{ \lower .75ex \hbox{$\sim$} \llap{\raise .27ex \hbox{$<$}} } 
\def\be{\begin{equation}} 
\def\ee{\end{equation}}

\def\mpl{M_{Pl}}

\begin{document} 
 
\title{Folded Inflation,  Primordial Tensors, and the Running of the  Scalar Spectral Index} 
 
\author{Richard Easther}
 
%\affiliation{~}
\affiliation{Department of Physics, Yale University, New Haven  CT 06520, USA}
 
\begin{abstract} 
I discuss {\em folded inflation\/}, an inflationary model embedded in a multi-dimensional  scalar potential, such as  the stringy landscape.   During folded inflation, the field point evolves along a path that turns several corners in the potential.  Folded inflation  can lead to  a  relatively large tensor contribution to the Cosmic Microwave Background, while keeping all fields smaller than the Planck scale.   I conjecture that  if  folded inflation generates a significant primordial tensor  amplitude, this will generically be associated with  non-trivial scale dependence in the spectrum of primordial scalar perturbations.   

\end{abstract} 
 
\maketitle 

\section{Introduction}

The spectrum of primordial perturbations is unlikely to be strictly scale invariant or, equivalently, the scalar spectral index, $n_R$,  is unlikely to be exactly unity. Conversely, it is often assumed that $n_R$  is itself independent of the wavenumber $k$.  However, the first year's data from the Wilkinson Microwave Anisotropy Probe [WMAP]  yielded two tantalizing hints of scale dependence in the perturbation spectrum: a  lack of power at small $l$ and, when combined with other datasets, a weak preference for $\frac{d n_R}{d\ln{k}} \neq 0$   \cite{Peiris:2003ff}.    Any hint that the primordial universe is {\em non-vanilla\/} \cite{Easther:2003fy} is of crucial importance, since this would  constrain both inflation and competing scenarios of the early universe. 

The  evidence for  scale dependence  is tentative. At small $l$ the data is accurate, but the lack of power may be due  to cosmic  variance \cite{Efstathiou:2003hk}.   Conversely,   the apparent evidence for $\frac{d n_R}{d\ln{k}} \neq 0$ could easily be a statistical artifact but if $\frac{d n_R}{d\ln{k}}$ is close to the current central  value  we will soon know this with some certainty.   Even if both effects are real, they may both be manifestations of the same underlying physics, or  two different phenomena.   

It is often said that a substantial value for $\frac{d n_R}{d\ln{k}} $ is unlikely,  since it suggests a  significant departure from smoothness in the inflaton potential.  Cosmological measurements probe a small range in $k$-space, corresponding to around 10 $e$-foldings of inflation, and  putting a ``feature'' in the potential that affects scales inside this narrow window  requires tuning.   However, this paper presents a scenario where scale dependence is not only permissible,  but expected.  

My starting point  is Lyth's observation that at high energies the total excursion made by the inflaton exceeds the Planck scale \cite{Lyth:1996im}.  As Lyth points out, this is  worrying  if the inflationary potential has  a stringy or supergravity origin: in these models the potential  is steep when any field value exceeds the Planck scale.      If  the inflation  scale is low (relative to the GUT scale),  the inflaton evolves comparatively slowly, resolving the conundrum.   The   primordial tensor  spectrum rises with the energy scale, and Lyth argued that  if inflation is  consistent with supergravity the tensor contribution must be very small.    

Suppose inflation is driven by several distinct fields, but  only  one field  is typically evolving at any given time, so the overall inflationary trajectory follows a path with several ``corners''.  We dub this model {\em folded inflation\/}.    On the face of it, folded inflation is extremely contrived.  However,  in string theory the potential surface for the light scalar fields is now thought to be a complicated and rugged
landscape.   This is an unknown  multi-dimensional function with  as many as 500  scalar degrees of freedom  \cite{Susskind:2003kw}. The landscape has  an exponentially large number of extrema, and it also has an exponentially large number of {\em paths\/}.  Some small subset of these paths will be suitable candidates for  folded inflation.  During folded inflation  the individual fields remain sub-Planckian but the total change in all the fields may exceed the Planck scale.   The hope is  thus that of the huge number of different ``downhill'' paths within the landscape, at least one of them can drive a cosmologically acceptable period of inflation   without the need for additional tuning.

For folded inflation to work at high scales, Lyth's argument forces the presence of  several corners during the last 60 e-folds of inflation. These corners will usually be associated with significant scale dependence in the perturbation spectrum.  At a sufficiently high energy scale, any given 10 e-folding window in the spectrum will almost certainly  contain a corner-induced feature. A similar argument for a scale-dependent spectrum is presented in \cite{Adams:1997de,Adams:2001vc}. For folded inflation,  $\frac{d n_R}{d\ln{k}} \neq 0$  is thus natural if inflation occurs at  high energy densities.

In what follows,  I  review the connection between the evolution of the  inflaton and the energy scale, and summarize the  perturbation spectra in models with multiple fields.    I show that the  spacing between  corners in the inflationary trajectory is correlated with the tensor amplitude. If the tensors are readily detectable, folded inflation  leads  to  a significant value for  $\frac{dn_R}{d\ln{k}}$.  Without a clear description of the stringy landscape, the discussion in this paper is necessarily more qualitative than quantitative.    However, we   identify lines of enquiry that will sharpen our understanding of the inflationary phenomenology of the  overall stringy landscape.

\section{The Lyth Bound}

Lyth's argument is very simple, and we repeat it here  in the  notation of   \cite{Peiris:2003ff}.  The scalar (density)  and tensor perturbations for a single, slowly rolling field  are 
\begin{eqnarray}
\Delta_R^2 &=&    \frac{1}{24 \pi^2  \epsilon} \frac{V}{\mpl^4} \, ,\\ 
\Delta_h^2 &=&  \frac{2}{3 \pi^2} \frac{V}{\mpl^4}   \, , \\
r &=& \frac{ \Delta_h^2}{\Delta_R^2} = 16 \epsilon \, .
\end{eqnarray}
Here $\mpl = 2.43 \times 10^{18} \mbox{GeV} $ is the reduced Planck mass, and $\epsilon$ is the first term in the (potential)  slow roll expansion,
\begin{eqnarray}
\epsilon &=& \frac{\mpl^2}{2} \left( \frac{V'}{V} \right)^2 \, ,\\
\eta &=& \mpl^2 \frac{V''}{V} \, ,\\ 
\xi &=& \mpl^4 \frac{ V' V''' }{V^2} \, .
\end{eqnarray}
The scalar and tensor spectral indices are
\begin{eqnarray}
n_R &=& 1- 6 \epsilon + 2 \eta  \, ,   \label{nr}  \\
\frac{d n_R}{d \ln{k}} &=& 16 \epsilon \eta - 24 \epsilon^2 -2 \xi \,  ,   \label{dnr}\\
n_h &=& -2 \epsilon \, , 
\end{eqnarray}
where, as usual, scale independent spectra correspond to $n_R = 1$ and $n_h = 0$. The ratio of the tensor and scalar  spectra is correlated with the slope of the tensor spectrum, leading to the slow-roll consistency condition.

We know the amplitude of density perturbation spectrum accurately from WMAP,  
\begin{equation}
\Delta_R^2(k_0) = \left( \frac{5}{3}\right)^2 \frac{800 \pi^2}{T^2} A(k_0)
 \end{equation}
where  $T$ is the CMB temperature in $\mu K$, and $k_0$ is a fiducial wavenumber.  From the 1-Year WMAP data,
 \begin{equation}
\Delta_R^2(k_0)   =   2.95 \times 10^{-9} A_0  \approx 2 \times 10^{-9} \, .
 \end{equation}
Since this will be a qualitative discussion we can safely omit uncertainties.  Using the number of e-folds $N = \ln(a)$ as a time variable,  $d N / dt = \dot{a}/a = H$,  and in the slow-roll limit 
 \begin{equation}
 \frac{1}{\mpl} \left| \frac{d \phi}{dN} \right| = \mpl \left| \frac{V'}{V} \right| \, .
 \end{equation} 
The right hand side of this equation is proportional to $\sqrt{\epsilon}$ and thus $\sqrt{r}$, so
 \begin{equation}
 \frac{1}{\mpl} \frac{d\phi}{dN} = \sqrt{\frac{r}{8}}\, .
 \end{equation}
During  $\Delta N$ e-folds,  $\Delta  \phi$  is
 \begin{equation}
\frac{\Delta\phi}{\mpl} = \sqrt{\frac{r}{8}} \Delta N  \, .
\end{equation}
For quasi de Sitter expansion,  $\Delta N \sim  \Delta \ln k \sim  \Delta \ln l$, where $l$ labels the CMB multipole.  Consequently, $\Delta N  \approx 7.6 $ over  the  range of scales probed by the CMB up to $l_{max} = 2000$, whereas $\Delta N \sim 60$ over the physically accessible portion of the inflationary era.   

Finding that  $\Delta \phi  \gsim \mpl$ is a significant embarrassment if we also imagine that  the inflationary potential is embedded in a supergravity model, or the stringy landscape.  The potential  acquires significant corrections when $|\phi| \sim \mpl$,  ensuring that $V'/V$ is large, ruining the flatness needed to support inflation.   If $\Delta \phi  \gsim \mpl$, the field must  evolve into a region where the potential has a substantial slope, telling us that our assumptions are not mutually compatible.    

Lyth argued for  a comparatively low value for the tensor amplitude, in order to ensure that inflation is consistent with supergravity.    Lyth took the minimal detectable value of $r$ to be $0.07$, and deduced that tensors are probably forever undetectable.   Today we can be more optimistic  -- with heroic efforts  $r \sim 6 \times 10^{-4}$  \cite{Knox:2002pe} or better \cite{Seljak:2003pn}  might be possible.   For now, the observational bound on $r$ remains high -- $ r < 0.7$ in the 1-Year WMAP dataset \cite{Peiris:2003ff}.  If we assume that the maximum excursion allowed for the field is $\Delta \phi \sim \mpl$ and $\Delta N = 60$ then we have $r \lsim  0.002$, which is well beyond the sensitivity of Planck \cite{Kinney:1998md}, but may  be possible in the distant future.  

\section{Folded Inflation}

 The analysis in the previous section implicitly assumed that inflation is driven by a single field, but paths in the stringy landscape can be folded into several different directions, such that $\sum_i | \Delta \phi_i| > \mpl$ but  $|\Delta \phi_i| <  \mpl$. 

The full expression for  the perturbations produced by multiple scalar fields   is \cite{Sasaki:1995aw} 
\begin{equation}
   \Delta_R^2 =  \left(\frac{H}{2 \pi}\right)^2  \sum_a \left( \frac{ \partial   N}{\partial \phi^a} \right)^2  \label{deltar2}
\end{equation}
if all the fields have canonical kinetic terms. The index $a$ runs over all the fields.  The  tensor amplitude is unchanged, since it is a function of the overall density. The consistency condition is now an inequality,
\begin{equation}
r \leq \frac{1}{8} |n_h|  \, .  \label{rmulti}
\end{equation}
The inequality is saturated if the field is free to move in only one direction \cite{LiddleLyth}, possibly after a redefinition of the fields.  Consequently, tensor modes are most likely to be detectable in the CMB if the inflationary potential has a unique ``downhill direction''.    The spectral index can again be written in terms of derivatives of the potential. To lowest order in the slow roll  expansion, $n_R$ depends on a mixture of first and second derivatives, whereas third derivatives appear in $\frac{d n_R }{d\ln{k}}$.  

Looking at (\ref{deltar2})  we see why a  saddle point in a multi-field potential is often associated  with  a dramatic rise in the power spectrum   \cite{Garcia-Bellido:1996qt}.   Firstly, near a saddle, a small change in the field value in the ``downhill'' direction can  cause a large change in the number of e-folds, leading to a large value for the relevant derivative in (\ref{deltar2}). Secondly,  within some region around the saddle point there are two downhill directions, not one, leading to a further amplification of the density peturbation spectrum.  This phenomenon is seen in the final phase of {\em locked inflation\/}  \cite{Dvali:2003vv,Easther:2004qs}, where  a saddle in the potential leads to large perturbations over some range of $k$, with runaway black hole production in the post-inflationary universe. 

Consequently, for folded  inflation to produce a realistic cosmology, the functional form of  the corners in the potential must  ensure they do not lead to an overproduction of black holes. This is analagous to the tight constraints on the parameter range open to locked inflation \cite{Easther:2004qs}.   In principle,  one could  design a folded inflation potential with a corner that did not lead to a  detectable  $\frac{d n_R}{d \ln{k}}$ signal. However, this is an additional and unnecessary constraint as there is no strong phenomenological reason for stipulating that $\frac{d n_R}{d\ln{k}} =0$. 

The WMAP team's analysis of the inflationary implications of their results \cite{Peiris:2003ff} included an analysis of inflation driven by the single--field potential \cite{Adams:2001vc}
\begin{equation}
V(\phi) = \frac{m^2}{2} \phi^2 \left( 1 + c \tanh\left( \frac{\phi- \phi_s}{d} \right) \right)  \label{step}
\end{equation}
This is not a multi-field potential, but the best-fit values of its parameters  \cite{Peiris:2003ff}  lead to a spectrum that is clearly $k$-dependent. While $\epsilon$ remains small, the higher order slow roll parameters are large, with $|\xi| \sim 100$ near $\phi_s$, signalling both strong scale dependence and a breakdown in the slow-roll approximation. 

Given that $\xi$ can take on large values over a small range of field values in the single field case,  we expect that  a generic corner in a random folded potential would correspond to a local feature in the perturbation spectrum.  This is especially true if we are relying on purely combinatorial arguments to motivate the existence of a folded inflation potential  embedded in the stringy landscape. Since the observational data accommodates a locally large value of  $\xi$  (and its multi-field generalization) as well as $\frac{d n_R}{d \ln{k}}$, insisting that the spectrum does not change significantly as the field point passes through the corner would amount to an {\em ad  hoc\/}  and needless tuning.

While folded inflation provides a way out of Lyth's constraint on the inflation scale,  there is  a  correlation between the spacing of corners in the potential, and the tensor amplitude.  Using data from both the CMB and large scale structure, we can  probe the power spectrum over  $\Delta \ln{k} \sim 10$.  To ensure that the field point does not turn a corner  as these scales leave the horizon,  we need  $r < 0.08$.  

This  is a conservative  bound, since it assumes that the observable part of the spectrum is exactly matched to the period of inflation between the corners. Moreover, as the results for potentials like (\ref{step}) indicate, a localized feature in the potential leads to a broader feature in $k$-space. Also, any feature is smeared in $l$-space, since each  CMB multipole samples a range of $k$-values.  In fact, two corners  in the potential could overlap when viewed in $l$-space. If we guess that the {\em average\/} separation of corners is roughly $M_{pl}$ (or deduce the likely separation by studying the combinatorics of the landscape) we could estimate the likelihood that the spectrum contains overlapping corners, as a function of $r$.

A different scenario, assisted inflation  \cite{Liddle:1998jc} relies on many fields rolling simultaneously, which cooperate to produce inflation.  In the stringy landscape, we could imagine starting many fields with the same initial value and letting them roll toward the center of the hypercube  defined where all  fields are sub-Planckian.  There are two reasons why assisted inflation is unlikely to be realized within the stringy landscape. Cross-couplings between the fields tend to undermine assisted inflation      \cite{Kanti:1999vt}, and while all the fields  in the string landscape need not couple at tree level, we do expect each field to be coupled to some of the others. Secondly, assisted inflation  requires a large number of fields to be prepared in  roughly the same initial state,  effectively reducing  the dimensionality  of the landscape. In the case of folded inflation, we have proposed the existence of a special path, but the usual objection to doing so is undermined by the combinatorics of a multidimensional potential. Reducing the effective dimensionality of the landscape undercuts the strength of these combinatorial arguments, making it unlikely that assisted inflation can be set up in this context. However, assisted inflation often possesses a formal attractor solution \cite{Malik:1998gy}, so it can be written in terms of an effective field theory where only one field is evolving , suggesting that the inequality in (\ref{rmulti}) is saturated, providing a different mechanism by which multi-field models can evade Lyth's constraint on the tensor amplitude.

\section{Summary and Future Directions}

The above qualitative analysis strongly suggests that if a) the inflationary potential is embedded within the stringy landscape or any other theory that contains supergravity, and b) the tensor to scalar ratio in the CMB, $r\gsim 0.08$, then the scalar spectrum is likely to  exhibit non-trivial scale dependence.  This is effectively the multi-field generalization of Lyth's argument that $r$ is low if inflation is driven by a single field.

Conversely, we do not need $r$ to be large  in order to observe a non-trivial $\frac{dn_R}{d \ln k}$. For instance \cite{Kadota:2003tn} describes a model with significant scale invariance, associated with a multi-component field moving in the string moduli space, but at an energy scale low enough to ensure that there is no observable tensor contribution to the CMB.   This is a quantitative analysis of a particular potential and trajectory that realizes folded inflation.    Several other inflationary models previously discussed in the literature can be interpreted as folded models 
\cite{Polarski:1992dq,Contaldi:2003zv,Feng:2003zu}.   Burgess {\em et al.\/} \cite{Burgess:2004kv}  examine realistic inflationary models motivated by the KKLMT proposal  \cite{Kachru:2003sx}, finding a possible tensor signal and models with a running index. This calculation applies to a  specific model, but its  conclusions overlap with those obtained on much more general grounds here.

Even if the inflationary trajectory does contain corners,  there is no {\em guarantee\/} that this will have an observable impact on the spectrum.   Inflation can  last until the density is well below the GUT scale, so that the region of the spectrum corresponding to the corners in the trajectory lies far outside the present horizon. Secondly, while I argued that a generic corner will result in a  non-trivial $\frac{d n_R}{d \ln{k}}$,  this outcome is not guaranteed. Furthermore, if corners produce localized features  analogous to those associated with potentials like equation (\ref{step}), the resulting spectrum will not have a ``constant'' running.    In future work, I plan to examine  the possible types of corners that could arise in an arbitrary multi-dimensional potential, using a multi-fiield generalization of {\em Monte Carlo reconstruction\/} \cite{Easther:2002rw}.    Likewise, progress in understanding  string phenomenology will give a better understanding of how many folded trajectories  exist within the landscape.  
 
Since the stringy landscape can easily support many possible inflationary trajectories, there is almost inevitably an anthropic element to this discussion. However, by requiring the phenomenological parameters that describe our universe -- both at the astrophysical and paticle levels -- be mutually consistent,  it may still be possible to make qualitative predictions based on the overall properties of the landscape. For example, Arkani-Hamed and Dimopoulos \cite{Arkani-Hamed:2004fb} recently used landscape based arguments  to identify  particles  physics signals correlated with the absence of low-energy supersymmetry.  It will be interesting to examine  anthropic bounds on the scale-dependence of $n_R$. If a large running is natural, an upper limit on $\frac{d n_R}{d\ln{k}}$ should be saturated in our universe, using a variant of Weinberg's argument for a  non-zero cosmological constant \cite{Weinberg:dv}. Anthropic bounds on the  amplitude of the density fluctuations have been proposed \cite{Tegmark:1997in,Garriga:1999hu} and generalizations of these arguments should lead to  constraints on the scale dependence of $n_R$.

In this paper, I have argued that when inflation occurs  at high energies, a scale dependent spectrum is a natural result. This is in  contrast to the usual theoretical prejudice against  running in the scalar spectrum, and if this correlation is confirmed observationally it will provide circumstantial evidence for the existence of the stringy landscape.
\vfill
\section*{ Acknowledgments}

I  thank Justin Khoury,  Will Kinney,  Hiranya Peiris, and Koenraad Schalm for a number of useful discussions.  This work is supported in part by  the United States Department of Energy, grant
DE-FG02-92ER-40704.


\begin{thebibliography}{99}

%\cite{Peiris:2003ff}
\bibitem{Peiris:2003ff}
H.~V.~Peiris {\it et al.},
%``First year Wilkinson Microwave Anisotropy Probe (WMAP) observations:
%Implications for inflation,''
Astrophys.\ J.\ Suppl.\  {\bf 148}, 213 (2003)
[arXiv:astro-ph/0302225].
%%CITATION = ASTRO-PH 0302225;%%


%\cite{Easther:2003fy}
\bibitem{Easther:2003fy}
R.~Easther,
%``Do We Live in a Vanilla Universe? Theoretical Perspectives on WMAP,''
AIP Conf.\ Proc.\  {\bf 698}, 64 (2004)
[arXiv:astro-ph/0308160].
%%CITATION = ASTRO-PH 0308160;%%


%\cite{Efstathiou:2003hk}
\bibitem{Efstathiou:2003hk}
G.~Efstathiou,
%``Is the low CMB quadrupole a signature of spatial curvature?,''
Mon.\ Not.\ Roy.\ Astron.\ Soc.\  {\bf 343}, L95 (2003)
[arXiv:astro-ph/0303127].
%%CITATION = ASTRO-PH 0303127;%%


%\cite{Lyth:1996im}
\bibitem{Lyth:1996im}
D.~H.~Lyth,
%``What would we learn by detecting a gravitational wave signal in the  cosmic
%microwave background anisotropy?,''
Phys.\ Rev.\ Lett.\  {\bf 78}, 1861 (1997)
[arXiv:hep-ph/9606387].
%%CITATION = HEP-PH 9606387;%%


%\cite{Susskind:2003kw}
\bibitem{Susskind:2003kw}
L.~Susskind,
%``The anthropic landscape of string theory,''
arXiv:hep-th/0302219.
%%CITATION = HEP-TH 0302219;%%


%\cite{Adams:1997de}
\bibitem{Adams:1997de}
J.~A.~Adams, G.~G.~Ross and S.~Sarkar,
%``Multiple inflation,''
Nucl.\ Phys.\ B {\bf 503}, 405 (1997)
[arXiv:hep-ph/9704286].
%%CITATION = HEP-PH 9704286;%%


%\cite{Adams:2001vc}
\bibitem{Adams:2001vc}
J.~Adams, B.~Cresswell and R.~Easther,
%``Inflationary perturbations from a potential with a step,''
Phys.\ Rev.\ D {\bf 64}, 123514 (2001)
[arXiv:astro-ph/0102236].
%%CITATION = ASTRO-PH 0102236;%%



%\cite{Liddle:1998jc}
\bibitem{Liddle:1998jc}
A.~R.~Liddle, A.~Mazumdar and F.~E.~Schunck,
%``Assisted inflation,''
Phys.\ Rev.\ D {\bf 58}, 061301 (1998)
[arXiv:astro-ph/9804177].
%%CITATION = ASTRO-PH 9804177;%%

%\cite{Kanti:1999vt}
\bibitem{Kanti:1999vt}
P.~Kanti and K.~A.~Olive,
%``On the realization of assisted inflation,''
Phys.\ Rev.\ D {\bf 60}, 043502 (1999)
[arXiv:hep-ph/9903524].
%%CITATION = HEP-PH 9903524;%%


%\cite{Knox:2002pe}
\bibitem{Knox:2002pe}
L.~Knox and Y.~S.~Song,
%``A limit on the detectability of the energy scale of inflation,''
Phys.\ Rev.\ Lett.\  {\bf 89}, 011303 (2002)
[arXiv:astro-ph/0202286].
%%CITATION = ASTRO-PH 0202286;%%


%\cite{Seljak:2003pn}
\bibitem{Seljak:2003pn}
U.~Seljak and C.~M.~Hirata,
%``Gravitational lensing as a contaminant of the gravity wave signal in CMB,''
Phys.\ Rev.\ D {\bf 69}, 043005 (2004)
[arXiv:astro-ph/0310163].
%%CITATION = ASTRO-PH 0310163;%%


%\cite{Kinney:1998md}
\bibitem{Kinney:1998md}
W.~H.~Kinney,
%``Constraining inflation with cosmic microwave background polarization,''
Phys.\ Rev.\ D {\bf 58}, 123506 (1998)
[arXiv:astro-ph/9806259].
%%CITATION = ASTRO-PH 9806259;%%


%\cite{Sasaki:1995aw}
\bibitem{Sasaki:1995aw}
M.~Sasaki and E.~D.~Stewart,
%``A General analytic formula for the spectral index of the density
%perturbations produced during inflation,''
Prog.\ Theor.\ Phys.\  {\bf 95}, 71 (1996)
[arXiv:astro-ph/9507001].
%%CITATION = ASTRO-PH 9507001;%%

\bibitem{LiddleLyth}
A . Liddle and D. Lyth, {\it Cosmological Inflation and Large Scale Structure\/}, Cambridge University Press, 2000.


%\cite{Garcia-Bellido:1996qt}
\bibitem{Garcia-Bellido:1996qt}
J.~Garcia-Bellido, A.~D.~Linde and D.~Wands,
%``Density perturbations and black hole formation in hybrid inflation,''
Phys.\ Rev.\ D {\bf 54}, 6040 (1996)
[arXiv:astro-ph/9605094].
%%CITATION = ASTRO-PH 9605094;%%


%\cite{Dvali:2003vv}
\bibitem{Dvali:2003vv}
G.~Dvali and S.~Kachru,
%``New old inflation,''
arXiv:hep-th/0309095.
%%CITATION = HEP-TH 0309095;%%

%\cite{Easther:2004qs}
\bibitem{Easther:2004qs}
R.~Easther, J.~Khoury and K.~Schalm,
%``Tuning locked inflation: Supergravity versus phenomenology,''
JCAP {\bf 0406}, 006 (2004)
[arXiv:hep-th/0402218].
%%CITATION = HEP-TH 0402218;%%

%\cite{Malik:1998gy}
\bibitem{Malik:1998gy}
K.~A.~Malik and D.~Wands,
%``Dynamics of assisted inflation,''
Phys.\ Rev.\ D {\bf 59}, 123501 (1999)
[arXiv:astro-ph/9812204].
%%CITATION = ASTRO-PH 9812204;%%

%\cite{Kadota:2003tn}
\bibitem{Kadota:2003tn}
K.~Kadota and E.~D.~Stewart,
%``Inflation on moduli space and cosmic perturbations,''
JHEP {\bf 0312}, 008 (2003)
[arXiv:hep-ph/0311240].
%%CITATION = HEP-PH 0311240;%%

%\cite{Polarski:1992dq}
\bibitem{Polarski:1992dq}
D.~Polarski and A.~A.~Starobinsky,
%``Spectra of perturbations produced by double inflation with an intermediate
%matter dominated stage,''
Nucl.\ Phys.\ B {\bf 385}, 623 (1992).
%%CITATION = NUPHA,B385,623;%%

%\cite{Contaldi:2003zv}
\bibitem{Contaldi:2003zv}
C.~R.~Contaldi, M.~Peloso, L.~Kofman and A.~Linde,
%``Suppressing the lower Multipoles in the CMB Anisotropies,''
JCAP {\bf 0307}, 002 (2003)
[arXiv:astro-ph/0303636].
%%CITATION = ASTRO-PH 0303636;%%

%\cite{Feng:2003zu}
\bibitem{Feng:2003zu}
B.~Feng and X.~Zhang,
%``Double inflation and the low CMB quadrupole,''
Phys.\ Lett.\ B {\bf 570}, 145 (2003)
[arXiv:astro-ph/0305020].
%%CITATION = ASTRO-PH 0305020;%%

\bibitem{Burgess:2004kv}
C.~P.~Burgess, J.~M.~Cline, H.~Stoica and F.~Quevedo,
%``Inflation in realistic D-brane models,''
arXiv:hep-th/0403119.
%%CITATION = HEP-TH 0403119;%%

%\cite{Kachru:2003sx}
\bibitem{Kachru:2003sx}
S.~Kachru, R.~Kallosh, A.~Linde, J.~Maldacena, L.~McAllister and S.~P.~Trivedi,
%``Towards inflation in string theory,''
JCAP {\bf 0310}, 013 (2003)
[arXiv:hep-th/0308055].
%%CITATION = HEP-TH 0308055;%%


%\cite{Easther:2002rw}
\bibitem{Easther:2002rw}
R.~Easther and W.~H.~Kinney,
%``Monte Carlo reconstruction of the inflationary potential,''
Phys.\ Rev.\ D {\bf 67}, 043511 (2003)
[arXiv:astro-ph/0210345].
%%CITATION = ASTRO-PH 0210345;%%

%\cite{Arkani-Hamed:2004fb}
\bibitem{Arkani-Hamed:2004fb}
N.~Arkani-Hamed and S.~Dimopoulos,
%``Supersymmetric unification without low energy supersymmetry and signatures
%for fine-tuning at the LHC,''
arXiv:hep-th/0405159.
%%CITATION = HEP-TH 0405159;%%

%\cite{Weinberg:dv}
\bibitem{Weinberg:dv}
S.~Weinberg,
%``Anthropic Bound On The Cosmological Constant,''
Phys.\ Rev.\ Lett.\  {\bf 59}, 2607 (1987).
%%CITATION = PRLTA,59,2607;%%


%
\bibitem{Tegmark:1997in}
M.~Tegmark and M.~J.~Rees,
%``Why is the CMB fluctuation level 10^{-5}?,''
Astrophys.\ J.\  {\bf 499}, 526 (1998)
[arXiv:astro-ph/9709058].
%%CITATION = ASTRO-PH 9709058;%%

%\cite{Garriga:1999hu}
\bibitem{Garriga:1999hu}
J.~Garriga, M.~Livio and A.~Vilenkin,
%``The cosmological constant and the time of its dominance,''
Phys.\ Rev.\ D {\bf 61}, 023503 (2000)
[arXiv:astro-ph/9906210].
%%CITATION = ASTRO-PH 9906210;%%

\end{thebibliography}
\end{document}